\documentclass[pra,twocolumn]{revtex4}
\usepackage{amsfonts,amsmath,amssymb,float}
\usepackage{bm,dsfont}
\usepackage{color}
\usepackage{bbm}
\usepackage{longtable}
\usepackage[dvips]{epsfig}
\usepackage{amsmath,amssymb,lscape,float}
\usepackage{hyperref}
\usepackage{listings}

\usepackage{graphicx}
\usepackage{epstopdf}
\usepackage{dcolumn}
\usepackage{bm}
\usepackage{amsfonts,amsmath,amssymb}

\usepackage{bm,dsfont}
\usepackage{color}
\usepackage{bbm}
\usepackage{longtable}
\usepackage[dvips]{epsfig}
\usepackage{hyperref}
\usepackage{listings}
\usepackage{todonotes}

\newcommand{\bra}[1]{\langle #1|}

\newcommand{\ket}[1]{|#1 \rangle}

\begin{document}
\title{Quantum-brachistochrone approach to the conversion from $W$ to 
Greenberger-Horne-Zeilinger states for Rydberg-atom qubits}
	
\author{Julian K. Nauth and Vladimir M. Stojanovi\'c}
\affiliation{Institut f\"{u}r Angewandte Physik, Technical
University of Darmstadt, D-64289 Darmstadt, Germany}
\date{\today}
\begin{abstract}
Using the quantum-brachistochrone formalism, we address the problem of finding the fastest possible (time-optimal)
deterministic conversion between $W$ and Greenberger-Horne-Zeilinger (GHZ) states in a system of three identical 
and equidistant neutral atoms that are acted upon by four external laser pulses. Assuming that all four pulses are 
close to being resonant with the same internal (atomic) transition -- the one between the atomic ground state and a 
high-lying Rydberg state -- each atom can be treated as an effective two-level system ($gr$-type qubit). Starting 
from an effective system Hamiltonian, which is valid in the Rydberg-blockade regime and defined on a four-state manifold, 
we derive the quantum-brachistochrone equations pertaining to the fastest possible $W$-to-GHZ state conversion. By numerically 
solving these equations, we determine the time-dependent Rabi frequencies of external laser pulses that correspond 
to the time-optimal state conversion. In particular, we show that the shortest possible $W$-to-GHZ state-conversion 
time is given by $T_{\textrm{QB}}= 6.8\:\hbar/E$, where $E$ is the total laser-pulse energy used, this last time 
being significantly shorter than the state-conversion times previously found using a dynamical-symmetry-based approach 
[ $T_{\textrm{DS}}=(1.33-1.66)\:T_{\textrm{QB}}$ ].
\end{abstract}
	
\maketitle
\section{Introduction} \label{Intro}
The availability of fast, accurate protocols for the preparation of highly-entangled quantum states is one 
of the crucial prerequisites for the adoption of next-generation quantum technologies~\cite{Dowling+Milburn:03}.
In particular, maximally-entangled multiqubit states of $W$-~\cite{Duer+:00} and GHZ~\cite{Greenberger+Horne+Zeilinger:89} 
type are of both conceptual and practical importance in the realm of quantum-information processing (QIP)~\cite{NielsenChuangBook}. 
Owing to their already proven usefulness in QIP~\cite{Joo+:03,Zhu+:15}, a large number of schemes for the preparation 
of $W$~\cite{Li+Song:15,Kang+:16,Kang+SciRep:16,StojanovicPRL:20,Peng+:21,StojanovicPRA:21,Zheng++:22,Haase++:22,Zhang+:22} 
and GHZ states~\cite{Coelho+:09,Song+:17,Erhard+:18,Macri+:18,Zheng+:19,Nogueira+:21,Pachniak+Malinovskaya:21,Qiao+:22} 
in various physical platforms have been proposed in recent years. Among those platforms, one of the most promising ones 
from the standpoint of large-scale quantum computing and analog quantum simulation is based on ensembles of neutral atoms 
in Rydberg states~\cite{Saffman:16,Henriet+:20,Morgado+Whitlock:21,ShiREVIEW:22}. Recent years have also seen growing general 
interest in quantum-state engineering in systems of this type~\cite{Buchmann+:17,Ostmann+:17,Malinovskaya:17,Omran+:19,Zheng+:20,
Mukherjee+:20,Shi:20,Pachniak+Malinovskaya:21,Haase+:21}.

Apart from various schemes for the efficient preparation of $W$ or GHZ states, which usually entail product states 
as their starting point~\cite{Omran+:19}, the interconversion between those two maximally-entangled states constitutes 
another relevant problem of quantum-state engineering. What makes the idea of interconversion between the two states -- which, 
for example, in the three-qubit case represent the only two inequivalent classes of genuine tripartite 
entanglement~\cite{Duer+:00} -- particularly appealing is the apparent dissimiliarity as far as the character of entanglement 
in the two states is concerned~\cite{Zhu+:22}. For instance, three-qubit GHZ state has maximal distributed (essential three-way-) 
entanglement, while pairwise bipartite entanglements all vanish~\cite{Coffman+:00}. On the other hand, for its $W$ counterpart 
the essential three-way entanglement vanishes, while it has strong pairwise entanglements~\cite{HorodeckiRMP:09}. 

The pioneering attempt of carrying out the $W$-to-GHZ state conversion pertained to a photonic system and its character was
probabilistic~\cite{Walther+:05}. Following this initial investigation, another work with photons was reported~\cite{Cui+:16},
as well as a study devoted to a spin system~\cite{Kang+:19}. In the realm of neutral-atom systems, irreversible conversions 
of a $W$ state into a GHZ state were first proposed~\cite{Song+:13,Wang+:16}, followed by two proposals for the deterministic 
conversion between the two states in a laser-controlled system of three equidistant $gr$-type Rydberg-atom qubits~\cite{Zheng+:20,
Haase+:21}. The first among those proposals~\cite{Zheng+:20} utilized the method of shortcuts to adiabaticity, more 
precisely an inverse-engineering approach based on a Lewis-Riesenfeld-type dynamical invariant. The second 
one~\cite{Haase+:21} -- advanced by one of us and collaborators -- was based on a Lie-algebraic approach that, under the 
assumption of real-valued Rabi frequencies of external laser pulses, explicitly takes into account the underlying dynamical 
symmetry $su(2)\oplus su(2)\cong so(4)$ of the effective system Hamiltonian defined on a manifold of four states (a basis of 
the permutation-symmetric subspace of the three-qubit Hilbert space). Importantly, the latter approach was shown to outperform
the former in the sense of allowing the envisioned state conversion to be carried out up to five times faster (for the same 
total laser-pulse energy used), with a much simpler time dependence of the corresponding Rabi frequencies of laser pulses 
used.
	
In this paper, aiming to find the time-optimal $W$-to-GHZ state conversion in the same physical setting as Refs.~\cite{Zheng+:20} 
and \cite{Haase+:21} -- i.e. for a system of three equidistant neutral atoms interacting through van-der-Waals-type 
interaction -- we employ the quantum-brachistochrone (QB) formalism~\cite{Carlini+:06}. Inspired in part by the 
time-honoured brachistochrone problem in classical mechanics~\cite{Haws+Kiser:95} this formalism was proposed by Carlini and co-workers 
with the aim of finding the fastest possible quantum evolution from a given initial to a desired final state~\cite{Carlini+:06}. 
It allows one to find time-optimal control protocols under the assumption that the system Hamiltonian has a bounded 
norm, being at the same time restricted to a subspace of Hermitian operators. Subsequently, the formalism was generalized so as to 
treat the problem of finding time-optimal unitary operations~\cite{Carlini+:07} and utilized for solving realistic gate-optimization 
problems~\cite{Wang++:15}. It has also led to important insights into a wide range of other problems of quantum 
physics~\cite{Caneva++:09,delCampo+:13,Lam+:21}.

Using the QB formalism we derive a system of first-order ordinary differential equations connecting the physical 
variables of interest in the problem at hand (namely, three time-dependent Rabi frequencies of external laser pulses, the 
corresponding auxiliary variables that have the nature of Lagrange multipliers, and the projections of the relevant quantum 
state of the system on the four relevant basis states). We then solve the two-point boundary value problem that corresponds 
to the shortest possible $W$-to-GHZ state conversion numerically using the {\em shooting method} (see, e.g., Ref.~\cite{NRcBook}), 
where the computational burden involved is substantially alleviated by means of an unconventional, problem-specific parametrization 
of the initial conditions. In this manner, we demonstrate that the time-optimal $W$-to-GHZ state conversion is $25$-$40\:\%$ 
faster than the protocol based on the dynamical-symmetry approach of Ref.~\cite{Haase+:21} that requires real-valued Rabi frequencies.
We also show that the three complex-valued (time-dependent) Rabi frequencies corresponding to the shortest-possible state conversion 
have time-independent phases. Moreover, we show that two of those phases can be chosen freely, with only the third one being 
constrained by the chosen values of the first two.

The remainder of this paper is organized as follows. In Sec.~\ref{System} we introduce the Rydberg-atom system under
consideration and state its previously derived effective Hamiltonian. In Sec.~\ref{QBforWtoGHZ} the QB equations governing 
the time-optimal conversion of an initial $W$ state into its GHZ counterpart in the Rydberg-atom system under consideration 
are derived, followed by a specific parametrization of their initial conditions that facilitates their subsequent numerical 
solution. The numerical solution of the QB equations is also discussed. In Sec.~\ref{ResDiscuss} we present and discuss the 
obtained results for the time-dependent Rabi frequencies of external laser pulses that correspond to the time-optimal state 
conversion. We also compare the obtained minimal state-conversion time with the corresponding times found in Refs.~\cite{Zheng+:20} 
and \cite{Haase+:21}. Finally, we demonstrate the robustness of our state-conversion scheme to deviations from the obtained 
optimal time-dependent Rabi frequencies. Before closing, we summarize the paper and underscore our main conclusions in 
Sec.~\ref{SummConcl}. The essential details of the method employed to numerically solve the QB equations are briefly reviewed 
in Appendix~\ref{NumericalDetails}.

\section{System and its effective Hamiltonian} \label{System}
\begin{figure}[b!]
\includegraphics[width=8.6cm]{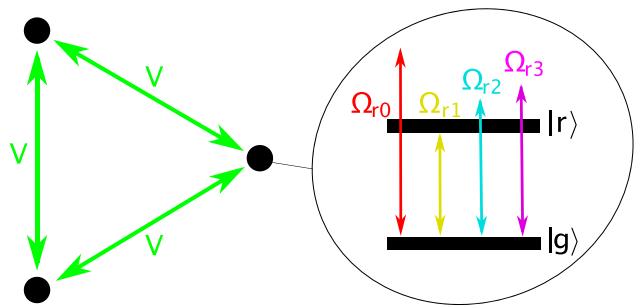} 
\caption{\label{fig:SystemIlustr}(Color online)Schematic of the system under consideration: Three identical and equidistant 
neutral atoms with the ground state $\ket{g}$ and a high-lying Rydberg state $\ket{r}$ ($gr$-type qubits) are acted upon by
four external laser pulses, all close to being resonant with the transition between the states  $\ket{g}$ and $\ket{r}$. 
$\hbar V$ is the magnitude of the van der Waals interaction.}
\end{figure}

In what follows, we consider a system of three identical and equidistant neutral atoms, e.g., of $^{87}$Rb [for an illustration, 
see Fig.~\ref{fig:SystemIlustr}]. All three atoms are subject to the same four external laser pulses, whose respective Rabi 
frequencies are denoted by $\Omega_{r0}$, $\Omega_{r1}$, $\Omega_{r2}$, and $\Omega_{r3}$. It is hereafter assumed that
the Rabi frequencies $\Omega_{r1}$, $\Omega_{r2}$, and $\Omega_{r3}$ are time dependent, while $\Omega_{r0}$ is time independent 
and envisioned to induce quadratic Stark shifts. 

The four laser pulses are assumed to be close to the resonance with the same internal atomic transition -- the one between the 
electronic ground state $|g\rangle$ and a highly-excited Rydberg state $|r\rangle$. Consequently, each atom can be treated as 
an effective two-level system -- a $gr$-type Rydberg-atom qubit~\cite{Morgado+Whitlock:21} where the states $|g\rangle$ and 
$|r\rangle$ play the role of logical $|0\rangle$ and $|1\rangle$ qubit states, respectively. The typical energy splitting of 
such qubits in frequency units is in the range $900 - 1500$\:THz (the actual energy splitting depends on the choice of atomic 
species and Rydberg states used), thus their manipulation requires either an ultraviolet laser or a combination of visible and 
infrared lasers in a ladder configuration. 

The Hamiltonian of the coupled atom-field system under consideration in the interaction picture is given by
\begin{eqnarray}\label{TotalHam}
H_{\textrm{I}}(t)/\hbar &=& \sum_{s=1}^3\sum_{n=0}^3\Omega_{rn}
(t)e^{-i(\delta_n+\Delta_n) t}\ket{r}_{ss}\bra{g} 
+ \text{H.c.} \nonumber \\
&+& \sum_{s<s'} V\ket{rr}_{ss'}\bra{rr} \:.
\end{eqnarray}
The first term describes the interaction between each of the three atoms (indexed by $s,s'=1,2,3$) and the four laser fields 
characterized by the Rabi frequencies $\Omega_{rn}$ ($n=0,\ldots,3$). The second one corresponds to the vdW atom-atom interaction 
with the pairwise interaction energy $\hbar V$. The detunings of the four laser pulses from the relevant internal ($g-r$) transition 
are split into two parts $\delta_n$ and $\Delta_n$ ($n=0,1,2,3$), in keeping with Ref.~\cite{Zheng+:20}. 

As demonstrated in Ref.~\cite{Zheng+:20}, through an appropriate choice of the detunings $\delta_n$ and $\Delta_n$ and under 
several conditions on other relevant parameters (interaction strength, laser-pulse duration, etc.)~\cite{Haase+:21}, the full 
system Hamiltonian of Eq.~\eqref{TotalHam} can be reduced via perturbation theory to an effective one defined on a four-state 
manifold. The relevant four three-qubit states are the three-atom ground state $\ket{ggg}$, the $W$ state 
\begin{equation}
\ket{W} = \frac{1}{\sqrt{3}}\:(\ket{rgg}+\ket{grg}+\ket{ggr}) \:, 
\end{equation}
the two-excitation Dicke state 
\begin{equation}
\ket{W'} = \frac{1}{\sqrt{3}}\:(\ket{rrg}+\ket{grr}+\ket{rgr}) \:, 
\end{equation}
and the state $\ket{rrr}$ with all three atoms occupying the Rydberg state. 

\begin{figure}[b!]
\includegraphics[width=8.6cm]{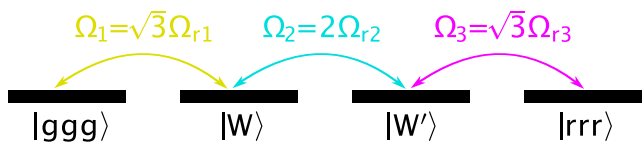} 
\caption{\label{EffectHamiltIllustr}(Color online)Pictorial illustration of the effective system Hamiltonian $H(t)$ defined
on a manifold of four states, where only pairs of adjacent states have nonzero coupling [cf. Eq.~\eqref{eq:HeffZheng}].}
\end{figure}

The effective Hamiltonian is given by~\cite{Zheng+:20}
\begin{eqnarray}
H(t)/\hbar &=& \Omega_1(t)\ket{ggg}\bra{W} +\Omega_{2}(t)\ket{W}\bra{W'} \nonumber\\ 
&+& \Omega_{3}(t)\ket{W'}\bra{rrr} +\text{H.c.} \label{eq:HeffZheng} \:,
\end{eqnarray}
where $\Omega_1(t)\equiv\sqrt{3}\:\Omega_{r1}(t)$, $\Omega_2(t)\equiv 2\:\Omega_{r2}(t)$, and $\Omega_3(t)\equiv\sqrt{3}\:\Omega_{r3}(t)$. 
In what follows, the time-dependent quantities $\Omega_n(t)$ ($n=1,2,3$), which differ from the original Rabi frequencies of external laser 
pulses only by constant prefactors, will also be referred to as Rabi frequencies. The structure of this last Hamiltonian, which is defined 
on a manifold of four states and has nonzero coupling only between adjacent ones, is illustrated in Fig.~\ref{EffectHamiltIllustr}.

Importantly, one of the conditions of validity of the effective Hamiltonian in Eq.~\eqref{eq:HeffZheng} is that $|V|T_{\textrm{int}}\gg 1$, 
where $T_{\textrm{int}}$ is the relevant laser-pulse duration~\cite{Haase+:21}. The last condition is equivalent to demanding that the 
interaction-induced energy shift $\hbar V$ is much larger than the Fourier-limited width of the laser pulses used, which precisely coincides 
with the definition of the Rydberg blockade (RB) regime. Thus, the above effective Hamiltonian is valid in the regime of primary interest 
for QIP with neutral atoms~\cite{Shi:18}.

Generally speaking, Hamiltonians that are defined on a manifold of four states and have nonzero {\em real-valued}
couplings only between adjacent states are characterized by the dynamical Lie algebra $su(2)\oplus su(2)\cong so(4)$ 
[this generalizes to analogous Hamiltonians defined on an $n$-state manifold, where the corresponding dynamical 
Lie algebra is $so(n)$]. Under the assumption of real-valued Rabi frequencies, this last dynamical symmetry was exploited 
in the context of $W$-to-GHZ state conversion in Ref.~\cite{Haase+:21}. In what follows, we refrain from the restriction 
to real-valued Rabi frequencies and aim to determine the fastest possible (time-optimal) deterministic conversion 
between an initial $W$ state and its GHZ counterpart.

Before embarking on the computation of the desired time-optimal $W$-to-GHZ state-conversion protocol, it is pertinent
to make the following, symmetry-related remark. Namely, it is important to note that the states $\ket{ggg}$, $\ket{W}$, 
$\ket{W'}$, and $\ket{rrr}$ form an orthonormal basis of a four-dimensional subspace of the (eigth-dimensional) three-qubit 
Hilbert space $\mathcal{H}\equiv(\mathbbm{C}^2)^{\otimes 3}$ that comprises the states invariant under an arbitrary permutation 
of qubits (i.e. invariant under the action of the symmetric group $S_3$); this four-dimensional subspace is usually referred 
to as the {\em symmetric sector} of the three-qubit Hilbert space~\cite{Ribeiro+Mosseri:11,Albertini+DAlessandro:18}. It is 
also worthwhile noting that for an arbitrary number of qubits -- including the special case $N=3$ of relevance here -- both 
$W$- and GHZ states are invariant under an arbitrary permutation of qubits. Because both the initial- and final states of our 
envisioned $W$-to-GHZ state-conversion process belong to the symmetric sector, the fact that the effective system Hamiltonian 
in Eq.~\eqref{eq:HeffZheng} involves the basis states of this particular subspace is pertinent from the symmetry standpoint.

\section{QB approach to the $W$-to-GHZ state conversion} \label{QBforWtoGHZ}
In the following, we make use of the QB formalism~\cite{Carlini+:06} to determine the time-dependent Rabi frequencies $\Omega_n(t)$ 
($n=1,2,3$) that pertain to the time-optimal conversion of an initial $W$ state $\vert\psi(t=0)\rangle\equiv\vert W\rangle$ of the 
system at hand to the GHZ-state $\vert\psi(t=T_{\textrm{QB}})\rangle\equiv(\vert ggg \rangle+e^{i\varphi}\vert rrr\rangle)/\sqrt{2}$.
Here $T_{\textrm{QB}}$ is the shortest possible state-conversion time, to be determined in what follows.

We first derive the QB equations pertaining to the time-optimal $W$-to-GHZ state conversion (Sec.~\ref{DeriveQBE}). We then discuss 
the numerical scheme that we employ to determine the solution to these equations (Sec.~\ref{NumSolveQBE}).

\subsection{Derivation of the QB equations for the time-optimal state conversion} \label{DeriveQBE}
We begin by representing the four three-qubit basis states $\vert ggg\rangle,\vert W\rangle,\vert W'\rangle$, and 
$\vert rrr\rangle$ by column vectors:
\begin{eqnarray}
\ket{ggg}&\mapsto&
\left(\begin{array}{c}
1\\
0\\
0\\
0
\end{array}\right),~~\ket{W} ~\mapsto~
\left(\begin{array}{c}
0\\
1\\
0\\
0
\end{array}\right),\nonumber\\
\ket{W'} &\mapsto&
\left(\begin{array}{c}
0\\
0\\
1\\
0
\end{array}\right),~~\ket{rrr}~\mapsto~
\left(\begin{array}{c}
0\\
0\\
0\\
1
\end{array}\right) \:.
\label{physicalstates}
\end{eqnarray}
This allows us to express the effective system Hamiltonian $H(t)$ [cf. Eq.~\eqref{eq:HeffZheng}] as
\begin{align}
H(t) = \begin{pmatrix}
0&\Omega_1(t)&0&0 \\
\Omega^{*}_1(t)&0&\Omega_2(t)&0 \\
0&\Omega^{*}_2(t)&0&\Omega_3(t) \\
0&0&\Omega^{*}_3(t)&0
\end{pmatrix}\:,
\end{align}
where, for the sake of notational convenience, we set $\hbar=1$ in the last equation; we will also keep
this convention throughout the following derivation of the QB equations.

The original QB formalism~\cite{Carlini+:06} allows one to find the Hamiltonian $H(t)$ whose corresponding 
dynamics enable the shortest possible evolution of the system under consideration from a given initial- to
a desired final state. That formalism is based on a quantum action of the form
\begin{eqnarray}\label{eq:QB_action}
&&S[H,\ket{\psi},\ket{\phi},\lambda] = \int\text{d}t\,\Bigg[ \frac{\sqrt{
\langle\dot{\psi}\ket{\dot{\psi}}-|\langle\dot{\psi}\ket{\psi}|^2}}{\Delta E} 
\notag\\&+& \sum_{j=1}^M\lambda_j f_j(H) +\left(i\langle\dot{\phi}\ket{\psi}
+\bra{\dot{\phi}} H\ket{\psi}+\text{c.c.}\right)\Bigg] \:.
\end{eqnarray}
Here $\Delta E\equiv\bra{\psi}H^2\ket{\psi}-\bra{\psi}H\ket{\psi}^2$ is the energy variance corresponding 
to the state $\ket{\psi}$  of the system and $\lambda$ the shorthand for the set of $M$ Lagrange multipliers 
$\lambda_j$ ($j=1,\ldots,M$); $\ket{\phi}$ is an auxiliary quantum state (a costate) that plays a role analogous
to that of $\lambda_j$. A variation of $H$ and $\ket{\psi}$ minimizes the Fubini-Study distance between the initial 
and final states, governed by the first term, and varying $\ket{\phi}$ in the third term ensures that $\ket{\psi}$ 
satisfies the Schr\"{o}dinger equation. The second term gives rise to $M$ additional, problem-specific constrains 
$f_j(H)=0$ by varying the Lagrange multipliers $\lambda_j$.

While in the original QB formalism $\text{Tr}[H^2(t)]$ is fixed at any time~\cite{Carlini+:06}, we weaken the 
latter restriction somewhat and fix only the total energy expended in the conversion process. In other words, 
the time-optimal state conversion is sought under the constraint of fixed total laser-pulse energy used, this 
last quantity being given by
\begin{align}\label{eq:fixed_E}
E = \int_0^T \text{d}t\,\sum_{n=1}^3 \vert\Omega_n(t)\vert^2 \:,
\end{align}
where $T$ is the -- as yet undetermined -- total evolution time of the system. The last requirement is taken into 
account by imposing the constraint
\begin{align}
f_1(H)=\frac{1}{2}\:\text{Tr} H^2 -\frac{E}{T} \:,
\end{align}
where $\text{Tr}\:H^2=2\sum_{n=1}^3 \vert\Omega_n\vert^2$, and choosing a constant Lagrange multiplier 
$\lambda_1$. 

Moreover, we must ensure that the form of the Hamiltonian is preserved during the entire process. In other
words, the time-dependent Hamiltonian of the system retains the form of Eq.~\eqref{eq:HeffZheng} throughout 
the conversion process. This is ensured by imposing the constraint $\sum_{j=2}^{M}
\lambda_j f_j(H)=\text{Tr} (\Lambda H)$, with
\begin{align}
\Lambda = \begin{pmatrix}
\lambda_\text{gg}&0&\lambda_\text{gW'}&\lambda_\text{gr} \\
0&\lambda_\text{WW}&0&\lambda_\text{Wr} \\
\lambda_\text{gW'}^*&0&\lambda_\text{W'W'}&0 \\
\lambda_\text{gr}^*&\lambda_\text{Wr}^*&0&\lambda_\text{rr}
\end{pmatrix}\:,
\end{align}
where each Lagrange multiplier in $\Lambda$ is time-dependent because $\text{Tr} (\Lambda H)=0$ ought to 
hold at all times. General constraints of this type have already been investigated in connection with 
quantum actions introduced in Ref.~\cite{Carlini+:06}. 

In line with Ref.~\cite{Carlini+:06}, we perform the variation of all variables in Eq.~\eqref{eq:QB_action} 
and arrive at the identities
\begin{align}\label{eq:F_psi}
F &= F\vert\psi\rangle\langle\psi\vert + \vert\psi\rangle\langle\psi\vert F 
- \vert\psi\rangle\langle\psi\vert \text{Tr} F\:,
\\\label{eq:F_dot_psi}
0&=(\dot{F}+i[H,F])\vert\psi\rangle\:,
\end{align}
where the operator $F$ involves the constraint functions $f_j(H)$ and is given by
\begin{align}\label{eq:F}
F = \sum_{j=1}^M \lambda_j \partial_H f_j(H) \:.
\end{align}
On account of the fact that $\lambda_1$ is a constant immaterial for our further discussion,
we set $\lambda_1=1$, which further yields $F=H + \Lambda$. In particular,
Eq.~\eqref{eq:F_dot_psi} implies that $F\vert\psi\rangle$ satisfies the Schr\"{o}dinger 
equation, which leads to the equation of motion~\cite{Carlini+:07}
\begin{align}\label{eq:F_dot}
\dot{F}=-i[H,F]\:.
\end{align}

We first make use of Eq.~\eqref{eq:F_psi} to derive the boundary values of $F$. By inserting 
$\vert\psi(0)\rangle$ and $\vert\psi(T)\rangle$ into this last equation, we readily obtain 
\begin{align}\label{eq:F_bound}
\notag
F(0) &= \begin{pmatrix}
0&\Omega_1(0)&0&0 \\
\Omega^{*}_1(0)&\lambda_\text{WW}(0)&\Omega_2(0)&\lambda_\text{Wr}(0) \\
0&\Omega^{*}_2(0)&0&0 \\
0&\lambda^{*}_\text{Wr}(0)&0&0
\end{pmatrix},
\\[0.5em]
F(T) &= \begin{pmatrix}
\lambda_\text{gg}(T)&\Omega_1(T)&e^{-i\varphi}\Omega^{*}_3(T)&\lambda_\text{gr}(T) \\
\Omega^{*}_1(T)&0&0&e^{-i\varphi}\Omega^{*}_1(T) \\
e^{i\varphi}\Omega_3(T)&0&0&\Omega_3(T) \\
\lambda^{*}_\text{gr}(T) &e^{i\varphi}\Omega_1(T)&\Omega^{*}_3(T)&\lambda_\text{rr}(T)
\end{pmatrix}.
\end{align}
The next step is to bring the equation of motion in Eq.~\eqref{eq:F_dot} to a more explicit form. By evaluating the 
diagonal elements of the matrix on the right-hand-side of this equation, one straightforwardly finds that these elements 
have the form of linear combinations $H_{jk}\Lambda_{jk}$, where either $H_{jk}$ or $\Lambda_{jk}$ is zero for each entry 
$(j,k)$. This leads to the conclusion that the time derivatives of $\lambda_\text{gg},\lambda_\text{WW},\lambda_\text{W'W'}$,
and $\lambda_\text{rr}$, which appear on the left-hand-side of Eq.~\eqref{eq:F_dot} are equal to zero. Therefore, 
$\lambda_\text{gg},\lambda_\text{WW},\lambda_\text{W'W'}$, and $\lambda_\text{rr}$ are constant. Moreover, by comparing 
$F(0)$ and $F(T)$ we conclude that $\lambda_\text{gg}=\lambda_\text{WW}=\lambda_\text{W'W'}=\lambda_\text{rr}=0$. 
At the same time, $\text{Tr} F^n$ turns out to be constant for each $n\in\mathbb{N}$. In particular, by making use of 
the fact that $\text{Tr}F^3(0)=\text{Tr}F^3(T)$ we find that $\lambda_\text{gr}(T)=ie^{-i\varphi}\vert\lambda_\text{gr}
(T)\vert$. On account of these last results, Eq.~\eqref{eq:F_dot} leads to the following system of (nonlinear) ordinary 
differential equations (ODEs): 
\begin{widetext}
\begin{align}\label{eq:F_dot_entries}
\partial_t \begin{pmatrix}
0&\Omega_1&\lambda_\text{gW'}&\lambda_\text{gr} \\
\Omega_1^*&0&\Omega_2&\lambda_\text{Wr} \\
\lambda_\text{gW'}^*&\Omega_2^*&0&\Omega_3 \\
\lambda_\text{gr}^*&\lambda_\text{Wr}^*&\Omega_3^*&0
\end{pmatrix}
=-i \begin{pmatrix}
0&-\Omega_2^*\lambda_\text{gW'}&-\Omega_3^*\lambda_\text{gr}&\Omega_1\lambda_\text{Wr}-\Omega_3\lambda_\text{gW'}
\\
\Omega_2\lambda_\text{gW'}^*&0&\Omega_1^*\lambda_\text{gW'}-\Omega_3^*\lambda_\text{Wr}&\Omega_1^*\lambda_\text{gr}
\\
\Omega_3\lambda_\text{gr}^*&\Omega_3\lambda_\text{Wr}^*-\Omega_1\lambda_\text{gW'}^*&0&\Omega_2^*\lambda_\text{Wr}
\\
\Omega_3^*\lambda_\text{gW'}^*-\Omega_1^*\lambda_\text{Wr}^*&-\Omega_1\lambda_\text{gr}^*&-\Omega_2\lambda_\text{Wr}^*&0
\end{pmatrix}\:.
\end{align}
\end{widetext}

The next step amounts to noticing that it is possible to reduce the number of equations of motion by half by proving 
that the complex phases $\phi_1,\phi_2,\phi_3,\phi_\text{Wr},\phi_\text{gr},\phi_\text{gW'}$ of the functions $\Omega_1,
\Omega_2,\Omega_3,\lambda_\text{Wr},\lambda_\text{gr},\lambda_\text{gW'}$ are constant in time. Assuming 
that those phases are time-independent, Eqs.~\eqref{eq:F_bound} and \eqref{eq:F_dot_entries} imply that the conditions
\begin{eqnarray}\label{eq:F_phases}
\phi_\text{Wr} &=& -\phi_1-\varphi\:,\nonumber\\
\phi_\text{gW'} &=& -\phi_3-\varphi\:, \nonumber\\ 
\phi_\text{gr} &=& -\varphi-\frac{\pi}{2}\:, \nonumber\\
\phi_2 &=& -\phi_1-\phi_3-\varphi-\frac{\pi}{2}\:,
\end{eqnarray}
ought to be satisfied; these restrictions imply that two out of three phases of the Rabi frequencies can be chosen arbitrarily 
(i.e. treated as free phases), while the third one is constrained by the chosen values of the first two. While the fulfillment 
of the conditions in Eq.~\eqref{eq:F_phases} -- i.e. the assumption of time-independent complex phases -- leads 
to one solution of the system in Eq.~\eqref{eq:F_dot_entries} (for each fixed choice of initial values), the fact 
that the latter is a system of first-order equations guarantees the uniqueness of this solution. In other words, if the complex 
phases of each entry of $F(0)$ are consistent with the restrictions in Eq.~\eqref{eq:F_phases}, then $F$ is uniquely determined 
by its initial value $F(0)$.

In the following, we consider $\phi_1$ and $\phi_3$ as free phases, while $\phi_2$ is determined from Eq.~\eqref{eq:F_phases} 
based on the values of $\phi_1$ and $\phi_3$. Given that the phases are found to be time-independent, we can reduce the problem
at hand to finding the moduli of $\Omega_n$ and $\lambda_k$. The equation of motion in Eq.~\eqref{eq:F_dot_entries} 
then reduces to
\begin{align}\label{eq:F_dot_real}
\partial_t \begin{pmatrix}
\vert\Omega_1\vert\\\vert\Omega_2\vert\\\vert\Omega_3\vert\\
\vert\lambda_\text{Wr}\vert\\\vert\lambda_\text{gr}\vert\\\vert\lambda_\text{gW'}\vert
\end{pmatrix}
&= \begin{pmatrix}
-\vert\Omega_2\vert \vert\lambda_\text{gW'}\vert\\
\vert\Omega_1\vert \vert\lambda_\text{gW'}\vert - \vert\Omega_3\vert \vert\lambda_\text{Wr}\vert\\
\vert\Omega_2\vert \vert\lambda_\text{Wr}\vert\\
-\vert\Omega_1\vert \vert\lambda_\text{gr}\vert\\
\vert\Omega_1\vert \vert\lambda_\text{Wr}\vert - \vert\Omega_3\vert \vert\lambda_\text{gW'}\vert\\
\vert\Omega_3\vert \vert\lambda_\text{gr}\vert
\end{pmatrix}.
\end{align}
Hence, we end up with the unknown initial values $\vert\Omega_1(0)\vert,\vert\Omega_2(0)\vert,\vert\lambda_\text{Wr}(0)\vert$ 
and the unknown final values $\vert\Omega_1(T)\vert,\vert\Omega_3(T)\vert,\vert\lambda_\text{gr}(T)\vert$. In addition, it is 
important to point out that the ODE system in Eq.~\eqref{eq:F_dot_real} does not depend on the phase $\varphi$ characterizing 
the GHZ state. This parameter can simply be determined by adjusting $\phi_2$, based on the constraints in Eq.~\eqref{eq:F_phases}. 
Moreover, it is worthwhile pointing out that the signs of the complex phases in Eq.~\eqref{eq:F_phases} were chosen in such a 
way that the moduli in Eq.~\eqref{eq:F_dot_real} remain positive during the entire state-conversion process.

Generally speaking, finding solutions of QB equations amounts to solving a two-point boundary value problem (BVP) in time. 
We now demonstrate that in this particular problem -- finding the time-optimal $W$-to-GHZ state conversion in the system at
hand -- the relevant BVP can be reduced to the one that involves only two unknown initial values that are also bounded. 
To this end, we first derive an inequality to bound $\vert\Omega_1(0)\vert$ and $\vert\Omega_2(0)\vert$. 
From Eq.~\eqref{eq:F_dot_real} it follows that
\begin{eqnarray}
\frac{E}{T} &=& \vert\Omega_1\vert^2+\vert\Omega_2\vert^2+\vert\Omega_3\vert^2 \:,\nonumber\\
\frac{E_\Lambda}{T} &=& \vert\lambda_\text{gW'}\vert^2+\vert\lambda_\text{gr}\vert^2+\vert
\lambda_\text{Wr}\vert^2 \:,
\end{eqnarray}
are both time-independent quantities. By evaluating both of these quantities at $t=0$ and $t=T$, we obtain
\begin{eqnarray}\label{eq:F_bound_real}
\frac{E}{T} &=& \vert\Omega_1(0)\vert^2+\vert\Omega_2(0)\vert^2 \:,\nonumber\\
\frac{E}{T} &=& \vert\Omega_1(T)\vert^2+\vert\Omega_3(T)\vert^2 \:,\nonumber\\
\frac{E_\Lambda}{T} &=& \vert\lambda_\text{Wr}(0)\vert^2\:,\nonumber\\
\frac{E_\Lambda}{T} &=& \vert\lambda_\text{gW'}(T)\vert^2+
\vert\lambda_\text{gr}(T)\vert^2+\vert\lambda_\text{Wr}(T)\vert^2\:.
\end{eqnarray}
The constraints in Eq.~\eqref{eq:F_bound} imply that $\vert\lambda_\text{Wr}(T)\vert=\vert\Omega_1(T)\vert$ and 
$\vert\lambda_\text{gW'}(T)\vert=\vert\Omega_3(T)\vert$, which -- when inserted in Eq.~\eqref{eq:F_bound_real} -- yields
\begin{align}
\frac{E_\Lambda}{T} =
\vert\lambda_\text{gr}(T)\vert^2+\frac{E}{T} \:.
\end{align}
The last equation leads to the conclusion that $E_\Lambda\geq E$ and, accordingly, 
$\vert\Omega_1(0)\vert^2+\vert\Omega_2(0)\vert^2\leq\vert\lambda_\text{Wr}(0)\vert^2$. 

The last conclusion enables us to bound the domain of the remaining initial values.
To this end, we eliminate $\vert\lambda_\text{Wr}(0)\vert$ by scaling the ODE system,
given by Eq.~\eqref{eq:F_dot_real}. We also define
\begin{eqnarray}\label{eq:scaling}
u_n &=& \frac{\vert\Omega_n\vert}{\vert\lambda_\text{Wr}(0)\vert}\quad(\:n= 1,2,3\:)\:,\nonumber\\
w_k &=&  \frac{\vert\lambda_k\vert}{\vert\lambda_\text{Wr}(0)\vert}\quad(\:k= \text{Wr},\text{gr}, 
\text{gW'}\:)\:,\\
\xi &=& \vert\lambda_\text{Wr}(0)\vert t \nonumber\:,
\end{eqnarray}
and the scaled (dimensionless) process time $\Xi= \vert\lambda_\text{Wr}(0)\vert T$, such that the ODE system 
is invariant under the transformation $\vert\Omega_n\vert\mapsto u_n$, $\vert\lambda_k\vert \mapsto w_k$, and 
$t\mapsto\xi$. It should be kept in mind that the new variables ought to fulfill the initial conditions
\begin{equation}
u_3(0)=0\:,\:w_\text{gr}(0)=0\:,\:w_\text{gW'}(0)=0 \:,
\end{equation}
as well as the final conditions
\begin{equation}\label{eq:uw_final}
u_2(T)=0\:,\:w_\text{Wr}(T)=u_1(T)\:,\:w_\text{gW'}(T)=u_3(T)\:,
\end{equation}
which can be derived from Eq.~\eqref{eq:F_bound}.

Owing to the restriction $\vert\Omega_1(0)\vert^2+\vert\Omega_2(0)\vert^2\leq \vert\lambda_\text{Wr}
(0)\vert^2$, we know that the point $[u_1(0),u_2(0)]^\text{T}$ lies within the unit circle, a bounded domain. 
It is thus pertinent to introduce polar coordinates $(u,\phi_u)$, in which $u_1(0)=u\cos\phi_u$, $u_2(0)=u\sin
\phi_u$, and the following identity holds true: 
\begin{equation}\label{EqUsq}
u^2 = \frac{|\Omega_1(0)|^2+|\Omega_2(0)|^2}
{\vert\lambda_\text{Wr}(0)\vert^2}\:.
\end{equation}
In this manner, we have reduced the unknown initial values to the values of the polar
radius $u$ and the polar angle $\phi_u$, which are both bounded, i.e.
\begin{equation}\label{eq:u,phiu}
u\in[0,1]\quad,\quad\phi_u\in[0,2\pi]\:.
\end{equation}

Using the column-vector representation of Eq.~\eqref{physicalstates} we obtain the time-dependent 
Schr\"{o}dinger equation 
\begin{align}\label{eq:Schroedinger}
\partial_t\begin{pmatrix}
\psi_\text{g}\\\psi_\text{W}\\\psi_\text{W'}\\\psi_\text{r}
\end{pmatrix}
=-i\begin{pmatrix}
\Omega_1\psi_\text{W} \\
\Omega_1^*\psi_\text{g}+\Omega_2\psi_\text{W'} \\
\Omega_2^*\psi_\text{W}+\Omega_3\psi_\text{r} \\
\Omega_3^*\psi_\text{W'}
\end{pmatrix}\:,
\end{align}
which governs the time evolution of the state
\begin{eqnarray} \label{statePsi}
|\psi(t)\rangle &=& \psi_\text{g}(t)|ggg\rangle + \psi_\text{W}(t)|W\rangle \nonumber \\
&+& \psi_\text{W'}(t)|W'\rangle + \psi_\text{r}(t)|rrr\rangle 
\end{eqnarray}
of the system. Once again, we simplify the ODE system resulting from Eq.~\eqref{eq:Schroedinger} by investigating, whether 
the phases $\phi_\text{g},\phi_\text{W},\phi_\text{W'},\phi_\text{r}$ of the state components $\psi_\text{g},\psi_\text{W},
\psi_\text{W'},\psi_\text{r}$ are constant. The form of Eq.~\eqref{eq:Schroedinger} bears out the assumption, if these
phases satisfy the constraints
\begin{eqnarray}\label{eq:phases_psi}
\phi_\text{g}  &=& \phi_\text{W}+\phi_1-\frac{\pi}{2}\:,\nonumber\\
\phi_\text{W'} &=& \phi_\text{W}+\phi_1+\phi_3+\varphi\:,\\
\phi_\text{r}  &=& \phi_\text{W}+\phi_1+\varphi-\frac{\pi}{2}\:,\nonumber
\end{eqnarray}
where one of the phases can be chosen arbitrarily, e.g., $\phi_\text{W}$. We emphasize that $\phi_\text{g}$ and 
$\phi_\text{r}$ only differ by $\varphi$, which is of crucial importance for generating the desired GHZ state. 

By making use of the last conclusion about the phases of the state components $(\psi_\text{g},\psi_\text{W},\psi_\text{W'},
\psi_\text{r})$, and applying the transformation in Eq.~\eqref{eq:scaling}, we can reduce the above time-dependent Schr\"{o}dinger
equation to an equation of motion that involves only the moduli of $(\psi_\text{g},\psi_\text{W},\psi_\text{W'},\psi_\text{r})$:
\begin{align}\label{eq:Schroedinger_real}
\partial_\xi\begin{pmatrix}
\vert\psi_\text{g}\vert\\\vert\psi_\text{W}\vert\\\vert\psi_\text{W'}\vert\\\vert\psi_\text{r}\vert
\end{pmatrix}
=\begin{pmatrix}
u_1 \vert\psi_\text{W}\vert \\
-u_1 \vert\psi_\text{g}\vert- u_2 \vert\psi_\text{W'}\vert \\
u_2 \vert\psi_\text{W}\vert-u_3 \vert\psi_\text{r}\vert \\
u_3 \vert\psi_\text{W'}\vert
\end{pmatrix}\:.
\end{align}
The boundary conditions inherent to the $W$-to-GHZ state-conversion problem under consideration are given by 
$|\psi_\text{W}(0)|=1$, $|\psi_\text{g}(0)|=|\psi_\text{W'}(0)|=|\psi_\text{r}(0)|=0$, $|\psi_\text{g}(\Xi)|=|\psi_\text{r}
(\Xi)|=1/\sqrt{2}$, and $|\psi_\text{W}(\Xi)|=|\psi_\text{W'}(\Xi)|=0$. Once again, it should be stressed that $\varphi$ only 
occurs in the restrictions of Eq.~\eqref{eq:phases_psi}, whereas this phase does not appear in Eq.~\eqref{eq:Schroedinger_real}. 
Similar to what was done in Eq.~\eqref{eq:F_dot_real} above, the signs of the complex phases in Eq.~\eqref{eq:phases_psi} 
are determined from the requirement that the moduli of the state components in Eq.~\eqref{eq:Schroedinger_real} ought to
remain positive during the process.

\subsection{Numerical solution of the QB equations} \label{NumSolveQBE}
As already pointed out above, finding solutions of QB equations amounts to solving a two-point BVP in time. Yet, this 
BVP is of a rather unconventional type as its final endpoint -- which, e.g., in the problem at hand corresponds to the 
minimal state-conversion time -- is unknown, being itself subject to minimization. While even standard two-point BVPs 
are far more demanding numerically than initial-value problems~\cite{NRcBook}, this additional aspect of QB equations 
typically renders such equations rather difficult to solve numerically. This is the principal reason as to why only a 
handful of problems of this type have as yet been efficiently solved numerically and certain specialized approaches for 
the numerical treatment of QB equations have been proposed. For instance, in Ref.~\cite{Wang++:15} an idea was proposed to 
treat QB paths as geodesics on the constraining manifold and determine the solutions of the QB equations by solving a set 
of geodesic equations. Another approach was proposed more recently~\cite{Wang+Shi+Lan:21}, which is based on a 
generalization of the original QB variational principle~\cite{Carlini+:06} and also makes use of the relaxation
method~\cite{NRcBook} for solving the ensuing BVP.

In the problem under consideration, owing to the proposed parametrization of initial conditions by only 
two variables $u$ and $\phi_u$ with bounded domains [cf. Eq.~\eqref{eq:u,phiu}], the relevant QB equations (i.e. the corresponding 
two-point BVPs) can efficiently be solved using the shooting method~\cite{NRcBook}. In other words, the character of the 
initial conditions in this problem obviates the need to use the aforementioned specialized numerical schemes. In line with 
the general idea of the shooting method, the initial values have to be chosen such that the functions occurring in the ODE 
system given by Eqs.~\eqref{eq:F_dot_real} and \eqref{eq:Schroedinger_real} (Rabi frequencies, Lagrange multipliers, and state 
components) satisfy the corresponding final conditions at $\xi=\Xi$, where $\Xi$ itself is as yet undetermined.

Within the framework of the shooting method, the initial conditions are modified iteratively in such a way that in the end 
the boundary conditions are fulfilled~\cite{NRcBook}. For each value of $u$ and $\phi_u$, the ODE system is solved within 
a certain time interval $[0,\xi_\text{max}]$. Through a global minimization we determine the dimensionless 
time within this interval for which the functions in the ODE system have the smallest deviations from their imposed final 
conditions (cf. Appendix \ref{NumericalDetails}). If the obtained smallest deviations vanish (i.e. if the functions do satisfy the 
final conditions) then the determined time is the sought-after minimal (dimensionless) state-conversion time $\Xi_{\textrm{QB}}$.
If such time cannot be found within the interval $[0,\xi_\text{max}]$ for any choice of $u$ and $\phi_u$ then a successful 
state conversion is not possible and the upper bound $\xi_\text{max}$ of the interval has to be increased. In other words, 
by varying the interval width $\xi_\text{max}$ we can verify whether the smallest possible value for $\Xi_{\textrm{QB}}$ 
was indeed obtained (for more details, see Appendix~\ref{NumericalDetails}).

It remains to clarify how to choose the starting points for the initial values. To avoid non-global minima in the aforementioned 
minimization of the deviation from the boundary conditions, it is necessary to check the entire domain of the initial values. In 
multi-dimensional problems, the shooting method can thus be a time- and resource-consuming approach~\cite{Wang+Shi+Lan:21}. However, 
here the domain of initial values is two-dimensional and bounded, which renders the problem at hand significantly simpler than in 
the generic case. As a result, we can perform the minimization starting from various initial values inside this domain with a moderate 
computational effort. This numerical computation, performed independently using three different minimization methods (for details, 
see Appendix~\ref{NumericalDetails}), yields the following values for $u$ and $\phi_u$: $(u,\phi_u) = (0.957,0.311\pi)$ and 
$(u,\phi_u)=(0.957,1.689\pi)$. 

\section{Results and Discussion} \label{ResDiscuss}
In the following, the principal findings of the present work -- based on the numerical solution of the 
QB equations -- are presented and discussed. In Sec.~\ref{TimeDependRabi} we first present the central 
result of this article -- the minimal state-conversion time $T_{\textrm{QB}}$. This is followed by the 
obtained results for the time-dependent Rabi frequencies that enable the time-optimal state conversion, 
and for the GHZ-state fidelity. To put the obtained results in perspective, in Sec.~\ref{CompareWithDS}
we compare them with those found in the previously used dynamical-symmetry-based approach. Finally,
in Sec.~\ref{Robustness} we demonstrate the robustness of the obtained time-optimal state-conversion 
scheme to deviations from the time-dependent Rabi frequencies found using the QB formalism.

\subsection{Minimal state-conversion time and time-dependence of Rabi frequencies} \label{TimeDependRabi}
By inserting the obtained values for $(u,\phi_u)$ into the relevant ODE system [cf. Eqs.~\eqref{eq:F_dot_real} and 
\eqref{eq:Schroedinger_real}] we obtain $\Xi_{\textrm{QB}}=2.72$ for the scaled state-conversion time in the time-optimal 
case. From this last value, it is straightforward to obtain the minimal state-conversion time $T_{\textrm{QB}}$ in 
terms of the natural timescale $\hbar/E$ in the problem at hand (we reinstate $\hbar$ for this purpose).

We first recall that $\vert\lambda_\text{Wr}(0)\vert=\sqrt{E_\Lambda/(\hbar T_{\textrm{QB}})}$ and 
$\Xi_{\textrm{QB}}=\vert\lambda_\text{Wr}(0)\vert T_{\textrm{QB}}$ [as follows from Eqs.~\eqref{eq:F_bound_real}
and \eqref{eq:scaling}, respectively], which immediately implies that $T_{\textrm{QB}}= \hbar\Xi_{\textrm{QB}}^2/E_{\Lambda}$. 
From Eqs.~\eqref{eq:F_bound_real} and \eqref{EqUsq} it follows that $E = u^2 E_\Lambda$. By combining the last
two expressions, we find that the minimal state-conversion time is given by
\begin{equation}
T_{\textrm{QB}} = u^2\Xi_{\textrm{QB}}^2\:\frac{\hbar}{E} \:.
\end{equation}
By inserting the obtained numerical results for $u$ and $\Xi_{\textrm{QB}}$ into the last expression, 
we finally obtain $T_{\textrm{QB}}=\:6.8\,\hbar/E$, which represents the central result of this paper.

\begin{figure}[b!]
\centering
\includegraphics[width=0.5\textwidth]{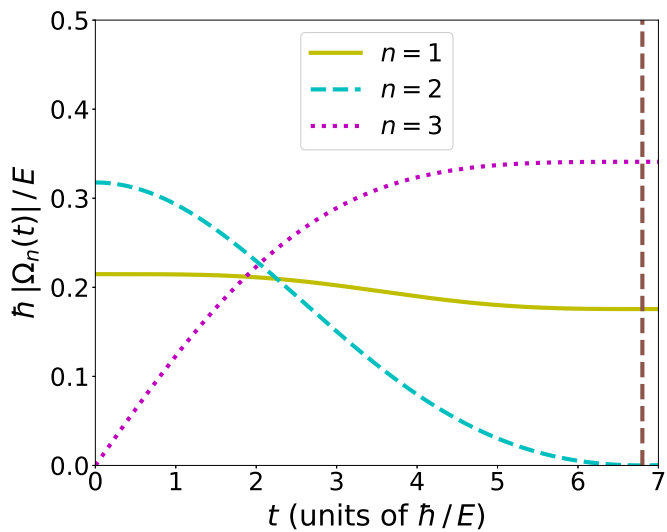}
\caption{\label{fig:Rabi frequencies}(Color online)Time evolution of the moduli of the complex-valued Rabi frequencies obtained 
numerically. The vertical dashed line marks the obtained minimal $W$-to-GHZ conversion time $T_{\textrm{QB}}=6.8\:\hbar/E$.}
\end{figure}

An equally important result of this paper pertains to the time dependence of the Rabi frequencies $\Omega_n(t)\equiv
\vert\Omega_n(t)\vert e^{i\phi_n}$ that corresponds to the shortest possible state-conversion process. By making 
use of the solution $(u_0,\phi_{u0})=(0.957,0.311\pi)$ for the underlying BVP we obtain the time dependence of 
the moduli $\vert\Omega_n(t)\vert$ of these Rabi frequencies 
depicted in Fig.~\ref{fig:Rabi frequencies}. It can easily be verified from this plot that the obtained results are 
consistent with the boundary conditions $\vert\Omega_3(0)\vert=0$ and $\vert\Omega_2(T_{\textrm{QB}})\vert=0$. Another 
interesting feature of the obtained results is that $\vert\Omega_3(t)\vert$ reaches a constant finite value at 
$t=T_{\textrm{QB}}$. The modulus $\vert\Omega_1(t)\vert$ of the third Rabi frequency does not vary appreciably 
during the process.

In view of the commonly occurring time-reversal-symmetric (or antisymmetric) solutions to quantum-control problems, 
it is pertinent to provide a comment as to why the Rabi frequencies in the problem at hand cannot be expected to display 
such behavior. Namely, this is apparent from Eq.~\eqref{eq:F_bound}, where $F(0)=\pm F(T)$ would lead to $\vert\Omega_2(0)
\vert=\vert\lambda_\text{gr}(0)\vert=\vert\lambda_\text{gW'}(0)\vert=0$, whereas Eq.~\eqref{eq:F_dot_real} would imply that 
$\vert\Omega_2\vert=\vert\Omega_3\vert=\vert\lambda_\text{gW'}\vert=0$ are constant, which obviously does not constitute a
solution of the problem at hand. Generally speaking, time-reversal-symmetric solutions for time-optimal processes are only 
expected if the initial and final state are related by a symmetry of the Hamiltonian~\cite{Wang+Shi+Lan:21}.

\begin{figure}[t!]
\includegraphics[width=0.5\textwidth]{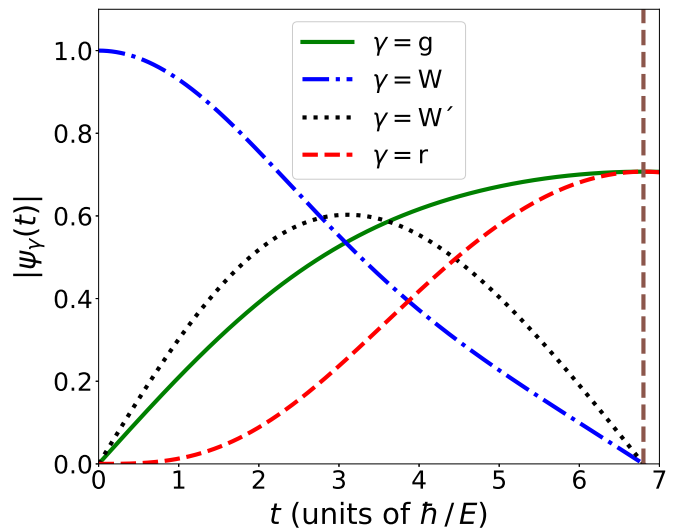}
\caption{\label{fig:State components}(Color online)Time dependence of the components $\vert\psi_\gamma(t)\vert$ of 
the state $\vert\psi(t)\rangle$ of the system obtained numerically. The vertical dashed line marks the obtained minimal 
$W$-to-GHZ conversion time $T_{\textrm{QB}}=6.8\:\hbar/E$.}
\end{figure}

\begin{figure}[b!]
\includegraphics[width=0.5\textwidth]{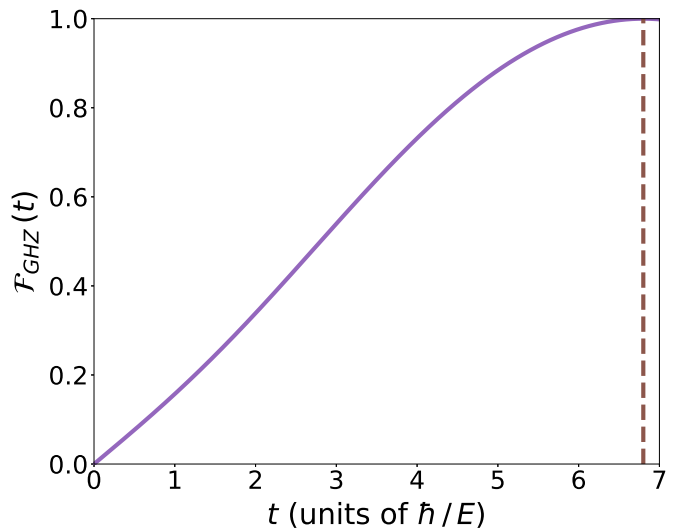}
\caption{\label{fig:Fidelity}(Color online)Time dependence of the GHZ-state fidelity $\mathcal{F}_\text{GHZ}(t)=\vert\langle
\text{GHZ}\vert\psi(t)\rangle\vert$ obtained numerically. The vertical dashed line marks the obtained minimal $W$-to-GHZ 
conversion time $T_{\textrm{QB}}=6.8\:\hbar/E$ at which the fidelity reaches unity.}
\end{figure}

While Fig.~\ref{fig:Rabi frequencies} only shows the moduli of the complex-valued time-dependent Rabi frequencies, 
it is pertinent to also comment at this point on their phases $\phi_1$, $\phi_2$, and $\phi_3$. As concluded in 
Sec.~\ref{QBforWtoGHZ}, these phases are constant (i.e. time-independent) and two out of three of them (e.g. $\phi_1$ 
and $\phi_3$) can take arbitrary values, while the remaining one is constrained by the chosen values of the first 
two [cf. Eq.~\eqref{eq:F_phases}]. Both of these last two properties of the phases inherent to $\Omega_n(t)$ ($n=1,2,3$) 
bode well for a potential experimental implementation of the laser pulses that correspond to the time-optimal state 
conversion. 

The central figure of merit characterizing the envisioned $W$-to-GHZ state-conversion process is the GHZ-state fidelity 
$\mathcal{F}_\text{GHZ}(t)$, which at time $t$ is given by the modulus of the overlap of the target (GHZ) state and the 
actual state $|\psi(t)\rangle$ of the system at that time $t$ [cf. Eq.~\eqref{statePsi}]. The time dependence of the four 
components ($\psi_\text{g}$, $\psi_\text{W}$, $\psi_\text{W'}$, $\psi_\text{r}$) of $|\psi(t)\rangle$ is illustrated in 
Fig.~\ref{fig:State components}, which -- among other things -- correctly reflects the fact that at the very beginning of 
the state-conversion process ($t=0$) we have that $|\psi_\text{W}|=1$, while at the end of this process ($t=T_{\textrm{QB}}$) 
we have $|\psi_\text{g}|=|\psi_\text{r}|=1/\sqrt{2}$.

Starting from the defining expression $\mathcal{F}_\text{GHZ}(t)=\vert\langle\text{GHZ}\vert\psi(t)\rangle\vert$, we 
straightforwardly find that the fidelity can be expressed as
\begin{equation}
\mathcal{F}_\text{GHZ}(t)= \frac{1}{\sqrt{2}} \left\vert e^{i\phi_\text{g}}\vert\psi_\text{g}(t)\vert+e^{-i\varphi}
e^{i\phi_\text{r}}\vert\psi_\text{r}(t)\vert\right\vert \:.
\end{equation}
On account of the fact that $\phi_\text{r}\equiv\phi_\text{g}+\varphi$, as implied by Eq.~\eqref{eq:phases_psi}, we
finally obtain that
\begin{equation}\label{GHZfidelity}
\mathcal{F}_\text{GHZ}(t)= \frac{\vert\psi_\text{g}(t)\vert+\vert\psi_\text{r}(t)\vert}{\sqrt{2}} \:.
\end{equation}
The GHZ-state fidelity -- obtained numerically based on the last expression -- is displayed in Fig.~\ref{fig:Fidelity},
from which it can be inferred that this fidelity shows monotonously increasing behavior and reaches unity at $t=T_{\textrm{QB}}$.

Importantly, the form of Eq.~\eqref{GHZfidelity}, which does not involve the phase $\varphi$ characterizing the target 
GHZ state, allows us to draw an important conclusion. Namely, the GHZ-state fidelity in the problem at hand does not depend 
on $\varphi$ at all. What makes this result plausible is the fact that only the phases $\phi_n$ of the complex-valued Rabi 
frequencies $\Omega_n(t)$ depend on $\varphi$, as can be inferred from the form of Eq.~\eqref{eq:F_phases}. Because these
three phases are time-independent it is plausible that they lead to a $\varphi$-independent GHZ-state fidelity at an arbitrary 
time $t$ during the state-conversion process ($0\le t\le T_{\textrm{QB}}$). This also seems to be consistent with the fact 
that the entanglement-related properties of GHZ states (e.g. the fact that they have maximal essential three-way entanglement, 
as quantified by the $3$-tangle~\cite{Coffman+:00}) also do not depend on $\varphi$.

\subsection{Comparison to the dynamical-symmetry-based approach} \label{CompareWithDS}
The $W$-to-GHZ state-conversion problem in the Rydberg-atom system under consideration has recently been addressed using a 
dynamical-symmetry-based approach~\cite{Haase+:21}. This approach allows one to carry out this conversion process up to 
five times faster than within the previously used shortcuts-to-adiabaticity approach~\cite{Zheng+:20}. 
It is thus pertinent to compare the state-conversion times $T_\text{DS}$ found using that approach with the minimal time 
$T_\text{QB}$ obtained here, where the comparison should be made under the assumption that the total laser-pulse energies 
used in both cases are the same.

Starting from the effective Hamiltonian in Eq.~\eqref{eq:HeffZheng}, the time dependence of the (real-valued) Rabi 
frequencies $\Omega_n(t)$ found in Ref.~\cite{Haase+:21} was shown to be given by
\begin{equation}
\Omega_n(t)= \frac{c_n}{T_\text{DS}(1-\tau)} 
f\left(\frac{t}{T_\text{DS}}\right) \:,
\end{equation}
where the function $f$ is defined as
\begin{equation}
f(x)=\begin{cases}
\frac{x}{\tau} &,\quad 0\leq x\leq \tau \\
1 &,\quad \tau\leq x\leq 1-\tau \\
\frac{1-x}{\tau} &,\quad 1-\tau\leq x\leq 1
\end{cases}\:,
\end{equation}
with $0\leq \tau\leq 1/3$. This last time dependence corresponds to functions $|\Omega_n(t)|$ that
grow from zero to a maximal value over the rise time $\tau\:T$, then remain constant during the time 
interval of the duration $T(1-2\tau)$, and, finally decay to zero during another interval of the duration 
$\tau\:T$. In the limiting case of vanishing rise/decay time ($\tau=0$), the corresponding pulse has 
a rectangular shape.

By inserting the last functional form of $\vert\Omega_n(t)\vert$ into the expression for the total
(fixed) laser-pulse energy [cf. Eq.~\eqref{eq:fixed_E}], we obtain
\begin{equation}
E =\hbar T_\text{DS} \sum_{n=1}^3 \left\vert \frac{c_n}{T_\text{DS}(1-\tau)}\right\vert^2 
\int_0^1 \text{d}x\, \vert f(x)\vert^2 \:.
\end{equation}
After a straightforward evaluation of the integral in the last equation, this finally leads to
\begin{equation}\label{EvsTds}
E=\frac{\hbar}{T_\text{DS}}\sum_{n=1}^3 \vert c_n\vert^2 
\frac{3-4\tau}{3(1-\tau)^2} \:.
\end{equation}
The coefficients $\vert c_n\vert$ in the last expression have the following values: 
$|c_1|=1.225$, $|c_2|=1.420$, and $|c_3|=2.352$. 

To be able to compare $T_\text{DS}$ determined from Eq.~\eqref{EvsTds} to the obtained minimal state-conversion 
time $T_\text{QB}$, it is sufficient to rearrange Eq.~\eqref{EvsTds} in order to express $T_\text{DS}$ in units of $\hbar/E$. 
In this manner, we find that $T_\text{DS}/T_{\textrm{QB}} = 1.66$ for $\tau=1/3$ and $T_\text{DS}/T_\text{QB} = 1.33$ for $\tau=0$. 
In other words, $T_\text{QB}$ is in the range between $0.6\:T_\text{DS}$ (for $\tau=1/3$) and $0.75\:T_\text{DS}$ (for $\tau=0$).
Thus, we arrive at the conclusion that the minimal state-conversion time $T_\text{QB}$ is $25-40\%$ shorter than the 
state-conversion times previously found using the dynamical-symmetry-based approach. Given that $T_\text{DS}$ is up to
$5$ times shorter than the state-conversion times obtained using shortcuts to adiabaticity~\cite{Zheng+:20}, we can also 
conclude that the minimal state-conversion time $T_\text{QB}$ is around $6.5$ times shorter than the latter times. 

Generally speaking, the capability of creating entanglement of a desired type on timescales significantly shorter than the 
coherence time of a quantum system is one of the prerequisites for QIP with that system. In particular, it has already been 
estimated that characteristic durations of $W$-to-GHZ state conversions using dynamical-symmetry-based approach are in the 
range $0.1-1\:\mu$s, which is $2-3$ orders of magnitude shorter than the typical radiative lifetimes of Rydberg states 
(around $100\:\mu$s for a state with the principal quantum number $n_q\sim 50$~\cite{GallagherBOOK}). The minimal conversion 
times $T_\text{QB}$ found here are even shorter, thus being significantly shorter than the relevant coherence times in
Rydberg-atom systems.

\subsection{Robustness of the state-conversion scheme against deviations from the optimal pulse shapes} \label{Robustness}
Generally speaking, in applications where a high degree of control over the dynamics of quantum systems is required 
it is of interest to be able to quantify the error pertaining to deviations from optimal-control solutions of various 
problems~\cite{Ho+:09,Negretti+:11,Stojanovic:19}. In other words, it is often of paramount importance to be able to design 
control pulses that -- while not being optimal -- yield an error (compared to the optimal solution) not larger than some 
pre-defined threshold value and are, at the same time, more amenable to an experimental implementation. 

\begin{figure}[b!]
\includegraphics[width=0.45\textwidth]{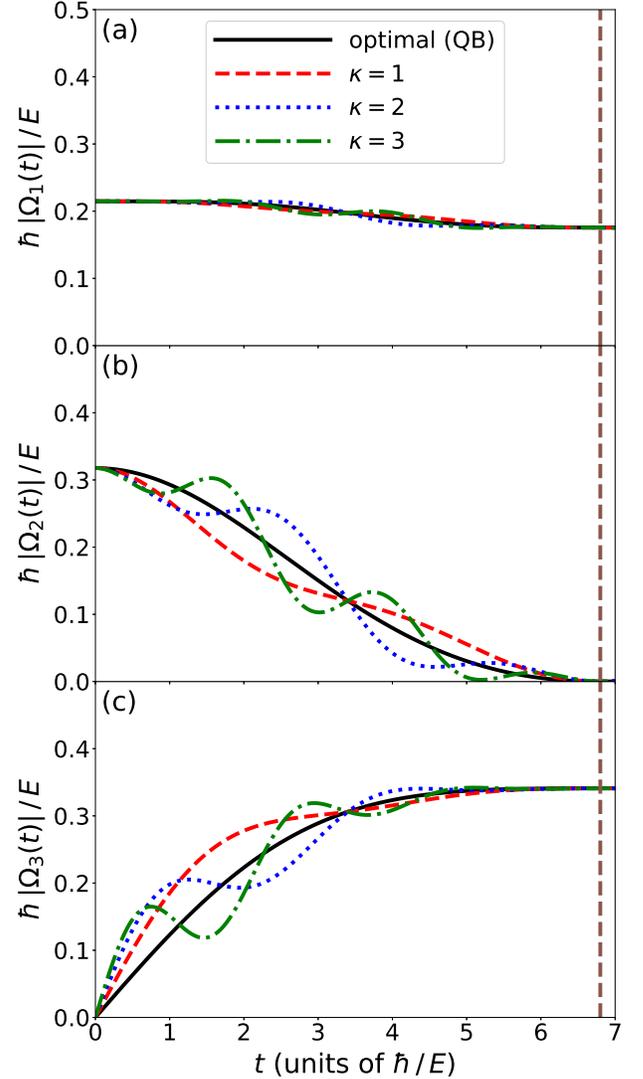}
\caption{\label{fig:RabiFreqDistort}(Color online)Time dependence of the distorted Rabi frequencies [cf. Eq.~\eqref{distortion}]
for (a) $n=1$, (b) $n=2$, and (c) $n=3$. The four curves for each value of $n$ correspond to the optimal Rabi frequency and
its three distorted counterparts (for $\kappa=1,2,3$). The chosen value of the parameter $t_n$ is equal to $0.1\:T_{\textrm{QB}}$.
The vertical dashed line marks the obtained minimal $W$-to-GHZ conversion time $T_{\textrm{QB}}=6.8\:\hbar/E$ at which the fidelity 
reaches unity.}
\end{figure}

In the $W$-to-GHZ state-conversion problem under consideration, using the QB formalism three time-dependent Rabi frequencies 
$\Omega_n(t)$ ($n=1,2,3$) have been determined that lead to the time-optimal state conversion (cf. Fig.~\ref{fig:Rabi frequencies}). 
In line with the above general considerations it is worthwhile to investigate the sensitivity of the state-conversion 
scheme at hand -- quantified by the GHZ-state fidelity $\mathcal{F}_\text{GHZ}$ at $t=T_{\textrm{QB}}$ -- with respect to 
deviations from the obtained optimal laser-pulse shapes. To this end, we consider the following form of (time-dependent) 
distortion from the optimal (QB) time dependence $\Omega_n(t)$ of the three Rabi frequencies~\cite{Negretti+:11}:
\begin{equation} \label{distortion}
\delta|\Omega_n(t)| = t_n \sin\left(2\pi\kappa\frac{t}{T}
\right)\:\frac{d}{dt}\:|\Omega_n(t)| \quad (\:n=1,2,3\:) \:.
\end{equation}
Here $\kappa$ is an integer-valued parameter that describes the rate of modulation of the optimal pulses, while the product 
of the prefactor $t_n$ -- which has units of time -- and the first derivative of the Rabi frequency $\Omega_n(t)$ represents 
the amplitude of distortion for $\Omega_n(t)$ $(n=1,2,3)$. This last form of distortion is very general, as it is capable of 
reproducing -- through an appropriate choice of the parameters $t_n$ and $\kappa$ -- almost any realistic pulse shape.

The form of the distorted moduli $|\Omega_n(t)|$ of the time-dependent Rabi frequencies is illustrated in Fig.~\ref{fig:RabiFreqDistort}. 
The much less pronounced distortion of $|\Omega_1(t)|$ [Fig.~\ref{fig:RabiFreqDistort}(a)] compared to $|\Omega_2(t)|$ and $|\Omega_3(t)|$ 
[Fig.~\ref{fig:RabiFreqDistort}(b) and Fig.~\ref{fig:RabiFreqDistort}(c), respectively] can straightforwardly be understood based on the 
fact that the original, time-optimal (i.e. obtained using the QB formalism) form of $|\Omega_1(t)|$ is characterized by a nearly time-independent
behavior (cf. Fig.~\ref{fig:Rabi frequencies}) and that the distortion in $|\Omega_n(t)|$ is -- by design -- proportional to its 
first derivative [cf. Eq.~\eqref{distortion}].

The GHZ-state fidelity $\mathcal{F}_\text{GHZ}(t=T_{\textrm{QB}})$ corresponding to the distorted Rabi frequencies is 
straightforwardly obtained once the state $|\psi(t=T_{\textrm{QB}})\rangle$ of the system at $t=T_{\textrm{QB}}$ is 
determined. This is accomplished by propagating the time-dependent Schr\"{o}dinger equation for the effective Hamiltonian 
of the system [cf. Eq.~\eqref{eq:HeffZheng}] in a numerically-exact fashion up to $t=T_{\textrm{QB}}$. 

\begin{figure}[b!]
\includegraphics[width=0.45\textwidth]{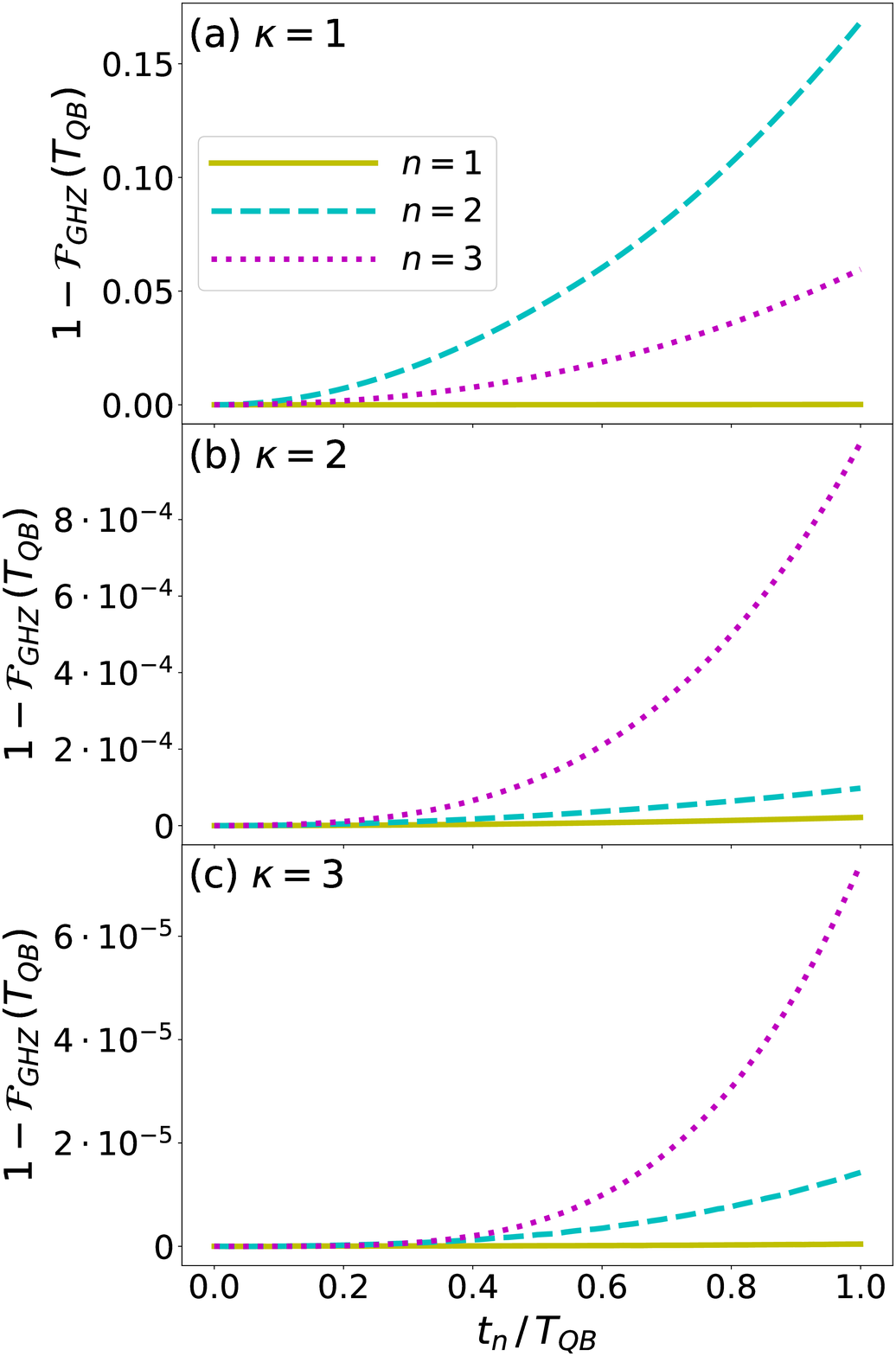}
\caption{\label{fig:Infidelity}(Color online)Deviation of the GHZ-state fidelity from unity (i.e. the infidelity),
when Rabi frequency $|\Omega_{n}|$ is distorted ($n=1,2,3$), at $t=T_{\textrm{QB}}$ for different values of the 
parameter $t_n$, shown for (a) $\kappa=1$, (b) $\kappa=2$, and (c) $\kappa=3$.}
\end{figure}

The obtained results for the deviation of the GHZ-state fidelity from unity at $t=T_{\textrm{QB}}$ [i.e. the infidelity
$1-\mathcal{F}_\text{GHZ}(t=T_{\textrm{QB}})$] as a result of distortion described by Eq.~\eqref{distortion} are displayed 
for different values of $t_n$ and $\kappa$ in Fig.~\ref{fig:Infidelity}. What can be inferred from these results is that
the reduction of the fidelity due to the assumed distortion of Rabi frequencies is quite small, being appreciable only
for $\kappa=1$ and extremely large values of $t_n$ [cf. Fig.~\ref{fig:Infidelity}(a)]. The extremely large values of 
$t_n$ -- those close to $T_{\textrm{QB}}$ -- correspond to rather drastic distortions from the original time-dependence 
of Rabi frequencies and are in fact not of practical relevance; realistic distortions are those corresponding to 
$t_n \lesssim T_{\textrm{QB}}/(2\pi\kappa)$, which for $\kappa=1$ amounts to $t_n \lesssim 0.15\:T_{\textrm{QB}}$.

Therefore, through the numerical evaluation of the resulting target-state fidelities we have demonstrated that our 
time-optimal $W$-to-GHZ state-conversion scheme is extremely robust to possible deviations from the optimal shape 
of the relevant Rabi frequencies of external lasers. This bodes well for future experimental implementations of the 
proposed state-conversion scheme.

\section{Summary and Conclusions} \label{SummConcl}
Using the quantum-brachistochrone formalism we have investigated the conversion of a $W$ state into 
its GHZ counterpart in a system that consists of three $gr$-type Rydberg-atom qubits acted upon by four external laser pulses. 
Starting from an effective system Hamiltonian, we derived the quantum-brachistochrone equations describing the time-optimal 
dynamical evolution from the original $W$ state to a GHZ state. We have solved  numerically the underlying two-point
boundary value problem using the shooting method and obtained the three time-dependent Rabi frequencies of external laser 
pulses that enable the desired, time-optimal state conversion. We have demonstrated that the minimal state-conversion time
$T_{\textrm{QB}}= 6.8\:\hbar/E$, where $E$ is the total laser-pulse energy used, is $25-40\:\%$ shorter than the state-conversion 
times recently obtained using a dynamical-symmetry-based approach with real-valued Rabi frequencies. In addition, we have also
shown that the proposed time-optimal state-conversion is extremely robust to deviations from the optimal laser-pulse 
shapes. 

Our work constitutes a highly nontrivial contribution to the growing body of work on quantum-state engineering in Rydberg-atom-based 
systems, one of the currently most promising platforms for large-scale quantum computing and analog quantum simulation. At the same time, 
the present work represents one of the very few examples to date of nontrivial quantum-control problems that have been fully solved within 
the quantum-brachistochrone framework. Our approach -- exploiting the form of the underlying equations to formulate a problem-specific 
parametrization of the initial conditions that drastically alleviates the computational burden that those equations entail -- could possibly 
provide guidelines for solving other nontrivial time-optimality-related problems in the realm of quantum control.

The present work is likely to motivate future studies as it can be generalized not only to other state-conversion problems in 
the Rydberg-atom system considered here, but also to systems belonging to several other physical platforms for quantum computing. 
An experimental implementation of the time-optimal $W$-to-GHZ state-conversion protocol obtained here is clearly called for.
\begin{acknowledgments}
V. M. S. acknowledges useful discussions with G. Alber and T. Haase. This research was supported 
by the Deutsche Forschungsgemeinschaft (DFG) -- SFB 1119 -- 236615297.
\end{acknowledgments}

\appendix

\section{Details of the numerical implementation}\label{NumericalDetails}
In the following, we provide the essential details of our numerical
implementation of the shooting method for solving the two-point BVP
of interest.

By making use of the proposed parametrization of initial conditions, we are now able to adjust the unknown initial 
values $\mathbf{x}\equiv(u,\phi_u)^\text{T}$, defined in Eq.~\eqref{eq:u,phiu}, and the scaled evolution time $\Xi$ 
with moderate computational effort. To guarantee that the final conditions of the variables $(u_1,u_2,u_3,w_\text{Wr},
w_\text{gr},w_\text{gW'})$ in Eq.~\eqref{eq:uw_final} and of the state components $(|\psi_\text{g}|,|\psi_\text{W}|,
|\psi_\text{W'}|,|\psi_\text{r}|)$ are fulfilled, we list them in the vector
\begin{equation}
\boldsymbol{d}_{\mathbf{x}}(\xi)\equiv
\begin{pmatrix}
u_2(\xi)\\
u_1(\xi)-w_\text{Wr}(\xi)\\
u_3(\xi)-w_\text{gW'}(\xi)\\
|\psi_\text{W}(\xi)|\\
|\psi_\text{W'}(\xi)|\\
|\psi_\text{g}(\xi)|-|\psi_\text{r}(\xi)|
\end{pmatrix}
\end{equation}
and minimize its Euclidean norm. To this end, it should be borne in mind that all functions appearing 
in the components of $\boldsymbol{d}_{\mathbf{x}}(\xi)$ depend on the initial values $\mathbf{x}$. 

\begin{figure}[t!]
\centering
\includegraphics[width=0.5\textwidth]{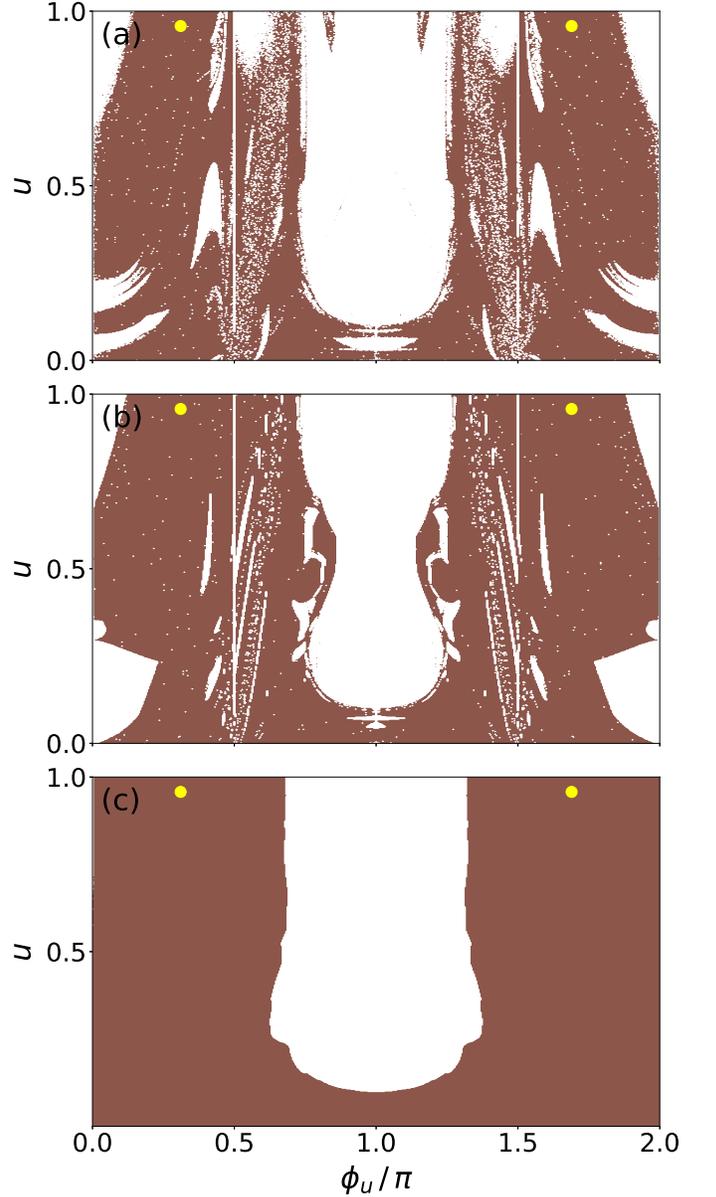}
\caption{\label{fig:Init_areas}(Color online)Illustration of the minimization of $D(\mathbf{x})$ with initial values 
$\mathbf{x}_0=(u_0,\phi_{u0})^\text{T}$ using (a) the Newton method, (b) the BFGS algorithm, and (c) the Nelder-Mead 
method. In the dark and the white areas, respectively, the computation terminates with an error $D(\mathbf{x})$ smaller 
and greater than $0.1\%$. The two dots mark points $(u_0,\phi_{u0})=(0.957,0.311\pi)$ and $(u_0,\phi_{u0})=(0.957,1.689\pi)$, 
where the global minima are reached.
}
\end{figure}

In the numerical implementation, we fix an upper bound $\xi_\text{max}$ of $\xi$. For given initial values $\mathbf{x}$, 
the ODE system in Eqs.~\eqref{eq:F_dot_real} and \eqref{eq:Schroedinger_real} can straightforwardly be solved using the 
\texttt{odeint} solver from the \texttt{scipy.integrate} package~\cite{odeint_scipy} of the SciPy library. We define the 
error as
\begin{align}\label{eq:D(x)}
D(\mathbf{x})=\min_{\xi\in[0,\xi_\text{max}]} \Vert\boldsymbol{d}_{\mathbf{x}}(\xi)\Vert_2 \:,
\end{align}
where $\Vert \mathbf{v}\Vert_2$ is the $l_2$-norm (i.e. Euclidean distance from the origin) of the vector $\mathbf{v}$.
The scaled evolution time $\Xi(\mathbf{x})\in[0,\xi_\text{max}]$ is the time that corresponds to the minimum in Eq.~\eqref{eq:D(x)}, 
i.e. $D(\mathbf{x})=\Vert\boldsymbol{d}_x\boldsymbol{(}\Xi(\mathbf{x})\boldsymbol{)}\Vert_2$. We minimize $D(\mathbf{x})$ with respect 
to $\mathbf{x}$ starting from $\mathbf{x}_0$. 

\begin{figure}[t!]
\centering
\includegraphics[width=0.5\textwidth]{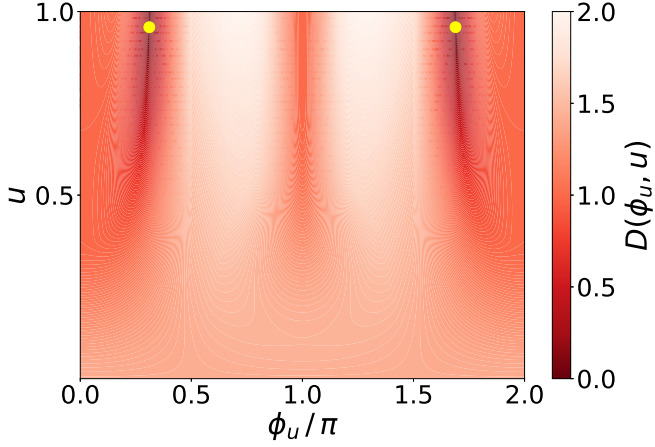}
\caption{(Color online)The error $D(\mathbf{x})\equiv D(u,\phi_{u})$ for initial values $(u,\phi_{u})$ 
without minimization with respect to $\mathbf{x}$. The two dots mark points $(u_0,\phi_{u0})=(0.957,0.311\pi)$ 
and $(u_0,\phi_{u0})=(0.957,1.689\pi)$, where the global minima are reached.}
\label{fig:Error directly}
\end{figure}

The minimization can be performed in several different ways. For instance, in Ref.~\cite{Wang+Shi+Lan:21} the Newton method was 
employed and the gradient of the function to be minimized was computed numerically. In keeping with this procedure, we compute 
the gradient of $D(\mathbf{x})$ by evaluating the derivatives of $\boldsymbol{d}_{\mathbf{x}}(\Xi(\mathbf{x}))$ according to 
\begin{eqnarray}
\partial_j D(\mathbf{x}) &=& \frac{\boldsymbol{d}_x\boldsymbol{(}\Xi(\mathbf{x})\boldsymbol{)}\cdot\partial_j\boldsymbol{d}_{\mathbf{x}}
\boldsymbol{(}\Xi(\mathbf{x})\boldsymbol{)}}{\Vert\boldsymbol{d}_{\mathbf{x}}\boldsymbol{(}\Xi(\mathbf{x})\boldsymbol{)}\Vert_2}\:,\\
\partial_j\boldsymbol{d}_{\mathbf{x}}\boldsymbol{(}\Xi(\mathbf{x})\boldsymbol{)} &=& \frac{\boldsymbol{d}_{\mathbf{x}+\epsilon \boldsymbol{e}_j}
\boldsymbol{(}\Xi(\mathbf{x}+\epsilon \boldsymbol{e}_j)\boldsymbol{)}-\boldsymbol{d}_{\mathbf{x}}\boldsymbol{(}\Xi(\mathbf{x})
\boldsymbol{)}}{\epsilon} \nonumber \:,
\end{eqnarray}
with $\partial_j$ ($j=1,2$) being a shorthand for the partial derivative $\partial/\partial_{x_j}$, where $x_1 \equiv u$ and 
$x_2\equiv \phi_u$; $\boldsymbol{e}_j$ denote the corresponding unit vectors. It is important to stress that accurate numerical 
evaluation of the derivatives in the last equation requires $\epsilon$ to be sufficiently small and the value $\epsilon=10^{-6}$ 
was used in the actual evaluation. It is also worthwhile mentioning that we adjust $\Xi$ in the course of our numerical procedure, 
in contrast to Ref.~\cite{Wang+Shi+Lan:21} where $\mathbf{x}$ is computed for fixed $\Xi$.

In addition to the Newton method, we solved the same problem independently using the Broyden-Fletcher-Goldfarb-Shanoo (BFGS) 
algorithm and the Nelder-Mead method~\cite{NRcBook}. We investigated various initial values $u_0=n_s/N$ and $\phi_{u0}=2\pi n_s/N$, 
where $n_s=0\ldots N$ and $N=500$ was chosen. Figure~\ref{fig:Init_areas} shows the areas in which a global minimum is found, 
that is, the error $D(\mathbf{x})$ drops below $0.1\%$. In these areas, the numerical computation terminates successfully at $(u_0,\phi_{u0})
=(0.957,0.311\pi)$ and $(u_0,\phi_{u0})=(0.957,1.689\pi)$, where $\Xi=2.72$ and $D(\mathbf{x}_0)=0.03\%$. Hence, all three minimization 
methods return the same values for global minima. For the sake of completeness, the error $D(\mathbf{x})$ itself (without minimization
over $\mathbf{x}$) is displayed in Fig.~\ref{fig:Error directly}.

To verify that $T_{\textrm{QB}}= 6.8\,\hbar/E$ indeed represents the shortest possible state-conversion time, we perform numerical
evaluation for different upper bounds $\xi_\text{max}$. Since $\mathbf{x}_0$ might vary, we compute the global minimum $\min_{\mathbf{x}} 
D(\mathbf{x})$ of the error $D(\mathbf{x})$ for each $\tau_\text{max}$ using different minimization methods. The corresponding state-conversion 
time $T$ turns out to be as large as possible for $\xi_\text{max}<2.72$ and stays constant for $\xi_\text{max}\geq 2.72$. Along with 
the observation that our calculations consistently show a strictly positive error until $T=T_{\textrm{QB}}=6.8\,\hbar/E$ is reached, 
this constitutes the evidence that $6.8\,\hbar/E$ is the minimal state-conversion time.

\end{document}